\documentstyle[aps,twocolumn]{revtex}
\newcommand{\be}{\begin{eqnarray}}
\newcommand{\ee}{\end{eqnarray}}

\renewcommand{\vec}{\bbox}

\begin{document}

%%%%%%%%%%%%%%%%%%%%%%%%%%%%%%%%%%%%%%%%%%%%%%%%%%%%%%%
% Title etc.
%%%%%%%%%%%%%%%%%%%%%%%%%%%%%%%%%%%%%%%%%%%%%%%%%%%%%%%
\title{Bose-Einstein Condensation of Atoms and Thermal Field 
Theory$^1$}

\author{Eric Braaten}

\address{Physics Department, Ohio State University, Columbus OH 43210, USA}

\date{\today}

\maketitle

\begin{abstract}
The Bose-Einstein condensation of atoms can be conveniently formulated as a
problem in thermal quantum field theory.  There are many properties 
of the equilibrium system and its
collective excitations that can be studied experimentally.  
The remarkable experimental control over all aspects of
the system make it an ideal testing ground for the methods of thermal
field theory.

\end{abstract}

\pacs{}

\narrowtext

%%%%%%%%%%%%%%%%%%%%%%%%%%%%%%%%%%%%%%%%%%%%%%%%%%%%%%%%%%%%%%%%%%%
\section{Introduction}
%%%%%%%%%%%%%%%%%%%%%%%%%%%%%%%%%%%%%%%%%%%%%%%%%%%%%%%%%%%%%%%%%%%

One of the most beautiful experimental developments in 
physics in this decade 
is the achievement of Bose-Einstein condensation in atomic gases.
This involves trapping a large number of atoms and then cooling 
them to such extremely low temperatures that they undergo a phase 
transition to a Bose-Einstein condensate.
This was first achieved in the summer of 1995 by a group at JILA
in Colorado using $^{87}$Rb atoms \cite{JILA}.  Bose-Einstein condensation
was soon achieved also at Rice using $^7$Li atoms \cite{Rice}
and at MIT using $^{23}$Na atoms \cite{MIT}.
There was a gap of a year and a half before any other laboratories
were able to achieve Bose-Einstein condensation, but by now the 
total number of groups that have succeeded is over a 
dozen.\footnotetext[1]{Invited talk presented at 
	Fifth International Workshop on Thermal Field Theories
	and Their Applications, Regensburg, August 1998.}

To get a feeling for what is involved in the Bose-Einstein 
condensation of trapped atoms, consider the simple case of 
$N$ identical noninteracting bosons trapped in an isotropic
harmonic oscillator potential $V(\vec r) = {1 \over 2} m \omega^2 r^2$
at temperature $T$.  These particles will form a cloud that is attracted 
to the deepest regions of the potential.  If $T$ is high enough that 
classical mechanics is applicable, the  radius $R$ of the cloud 
can be estimated using the equipartition theorem: 
\begin{equation}
{1 \over 2} m \omega^2 R^2 \; \sim  \; {p^2 \over 2 m}
 \; \sim  \; {3 \over 2} k T.
\label{equi}
\end{equation}
We find $R \sim (kT/m \omega^2)^{1/2}$, so that the radius of the cloud 
decreases as it is cooled.  At extremely low  temperatures, 
it is necessary to take into account the 
quantum mechanical nature of the states in the potential.
If $kT$ is small compared to the splitting $\hbar \omega$ between 
the energy levels of the harmonic oscillator, all the particles will be in
the ground state, which has a Gaussian wavefunction with width of 
order $\sqrt{\hbar/m \omega}$.  Thus the radius of the cloud
at zero temperature will be $R \sim \sqrt{\hbar/m \omega}$.
Einstein realized in the 1920's that the cooling of noninteracting 
bosons involves a phase transition.  The transition sets in when the 
wavefunctions of the particles begin to overlap, which is when
their deBroglie wavelengths become comparable to their separations:
\begin{equation}
\hbar/p \; \sim  \; \left(R^3 / N \right)^{1/3}.
\label{wave}
\end{equation}
Using (\ref{equi}) and (\ref{wave}), we find that  the radius 
of the cloud is $R \sim N^{1/6} \sqrt{\hbar/m \omega}$
and that the transition temperature is roughly
\begin{equation}
kT \; \sim \; N^{1/3} \hbar \omega
\label{Tc}
\end{equation}
At this temperature, atoms begin condensing into the ground state
and forming a dense core with radius $\sqrt{\hbar/m \omega}$
at the center of the cloud.  As the temperature is lowered further,
the core becomes more and more dense as more particles condense.
Meanwhile the thermal cloud becomes less and less dense and then
disappears altogether at $T = 0$. 

We can use the example of the ideal gas of bosons to get some idea 
of the scales involved in experiments on the Bose-Einstein 
condensation of atoms.  In a typical experiment, 
$N \sim 10^6$ atoms are trapped in a potential that can be 
approximated by a harmonic oscillator with angular frequency
$\omega \sim 10^3$ Hz.  From (\ref{Tc}),
the critical temperature for  Bose-Einstein condensation is 
roughly $kT_c \sim 10^{-10}$ eV, or $T_c \sim 10^{-6}$ K.  
We might expect the size of the condensate at $T=0$ to be
equal to that of the ground state wavefunction for a single atom,
which is roughly 1 micron.  However it is typically larger by
almost an order of magnitude.  This illustrates the important
fact that many of the properties of a Bose-Einstein condensate
of atoms are dramatically affected by the interactions between
the atoms.  This makes its behavior far more interesting than 
that of an ideal gas of noninteracting bosons.

%%%%%%%%%%%%%%%%%%%%%%%%%%%%%%%%%%%%%%%%%%%%%%%%%%%%%%%%%%%%%%%%%%%
\section{Atomic Physics}
%%%%%%%%%%%%%%%%%%%%%%%%%%%%%%%%%%%%%%%%%%%%%%%%%%%%%%%%%%%%%%%%%%%

There are a few aspects of atomic physics that one should be aware of in order
to appreciate the experiments on Bose-Einstein condensation.  I will describe
these aspects using the $^{87}$Rb atom as a specific example.

\subsection{Atomic Structure}

The $^{87}$Rb atom is an alkali atom with 36 electrons in closed shells with
total angular momentum 0 and one valence electron in an S-wave outer shell 
with spin $s = {1 \over 2}$.  It has a nucleus 
containing 37 protons and 50 neutrons with
total spin $i = {3 \over 2}$.  The total angular momentum vector of the
atom is ${\bf F} = {\bf I} + {\bf S}$, where ${\bf I}$ and ${\bf S}$ are the
angular momenta of the nucleus and the electrons.
The spin-spin interaction between the magnetic moments of 
the nucleus and the valence electron splits the ground
state of the atom into hyperfine states with total angular momentum 
$f=1$ and 2. The splitting between the two hyperfine energy levels is
tiny (the $f=1$ state for $^{87}$Rb is lower by about $3 \times 10^{-6}$ eV),
but this is enormous compared to the 
kinetic energies of atoms undergoing  Bose-Einstein condensation.
Thus it is the hyperfine states of the atom that are relevant in
Bose-Einstein condensation.

\subsection{Interactions between Atoms}

The interaction between two atoms can be described by a two-body potential 
$U({\bf r}_1 - \vec r_2)$ with a repulsive core, 
a short-range attractive well, and a long-range van der Waals tail.  
The potential for $^{87}$Rb contains 123 molecular bound states, 
but the splittings between the bound state energy levels are enormous
compared to the kinetic energies of atoms undergoing BEC.  At these extremely
low energies, the effects of the potential can be subsumed into 
a single number, the S-wave scattering length,
which is $a = 58$\AA \ for $^{87}$Rb.  
It is at first surprising that $a$ is an
order of magnitude larger than the range of the attractive well, which is
about 5\AA.  However, the scale of the scattering length is set not by the
short-range potential, but instead by the van der Waals tail 
$U(\vec r) \to \alpha/r^6$.  
The van der Waals length
$(m\alpha/\hbar^2)^{1/4}$ is defined by the balance between the 
kinetic energy and the van der Waals energy, and has the value
66 \AA \ for $^{87}$Rb.

\subsection{Trapping Potentials}

There are two kinds of traps that have been used to confine Bose-condensed
atoms:  magnetic traps and optical traps.  A magnetic trap relies on the
interaction between the magnetic moment of the valence electron 
and an applied magnetic field ${\bf B}(\vec r)$.  The relevant term in the
hamiltonian is
\begin{equation}
H_{\rm magnetic} = - {e \over m_e} {\bf S} \cdot {\bf B} (\vec r_e),
\end{equation}
where $\vec r_e$ and ${\bf S}$ 
are the position and spin of the valence electron.
Using first order perturbation theory, the Wigner-Eckhart theorem, and the
adiabatic approximation, the effects of the interaction can be represented by
the external potential
\begin{equation}
V_{\rm magnetic}(\vec r) 
= - m_f g_f {e \hbar \over m_e} | {\bf B} (\vec r)|,
\end{equation}
where $g_f = [f (f+1) + s (s+1) - i (i + 1)]/2f(f+1)$ and $m_f \hbar$ is the
projection of the total angular momentum ${\bf F}$ onto the local direction of
the magnetic field ${\bf B}(\vec r)$.  Those hyperfine states $|f, m_f \rangle$ for
which $m_f g_f$ is positive are attracted to the regions of strong field and
therefore can be confined in such a region.  In the case of $^{87}$Rb, these
state are $| 2, +2 \rangle$, $|2, +1 \rangle$, and $|1, -1 \rangle$.

Bose-condensed atoms can also be confined by an optical trap.  Such a trap
relies on the coupling of the electric dipole moment of the atom to an
external electric field ${\bf E}(\vec r, t)$ provided by laser beams.  The
relevant term in the hamiltonian is
\begin{equation}
H= -e (\vec r_e - \vec r_n) \cdot {\bf E}(\vec r_n, t),
\end{equation}
where $\vec r_e$ and $\vec r_n$ are the positions of the 
valence electron and the nucleus respectively.  
At first-order in perturbation theory, the interaction
averages to zero over time because the electric fields from the laser beams
are rapidly oscillating: 
$\langle {\bf E} (\vec r,t) \rangle_t = 0$.  
However, at second order in perturbation theory, the
time-average is nonzero and the effect of the interaction can be described by
a potential
\begin{equation}
V_{\rm optical} (\vec r) \, \propto \, 
\langle {\bf E}^2 (\vec r, t) \rangle_t .
\end{equation}
This optical potential can be used to trap any or all of the hyperfine states
$|f, m_f \rangle$.

%%%%%%%%%%%%%%%%%%%%%%%%%%%%%%%%%%%%%%%%%%%%%%%%%%%%%%%%%%%%%%%%%
\section{Theoretical Formulation}
%%%%%%%%%%%%%%%%%%%%%%%%%%%%%%%%%%%%%%%%%%%%%%%%%%%%%%%%%%%%%%%%%

A system consisting of a large number of low-energy atoms trapped in a 
potential can be conveniently described by a local quantum field theory.
We sketch the derivation of this formulation starting from the 
conventional description using ordinary quantum mechanics.

\subsection{Many-body Quantum Mechanics}

The problem of $N$ identical bosonic atoms trapped in an external potential
$V(\vec r)$ and interacting through an interatomic potential 
$U(\vec r_1 -\vec r_2)$ can of course be formulated using ordinary 
quantum mechanics.  The $N$ atoms are described by a Schr\"odinger wavefunction 
$\psi(\vec r_1, \vec r_2, \dots , \vec r_N ; t)$, 
which is a complex-valued function of $N$
coordinates $\vec r_i$ and the time $t$.  The bosonic nature of the atoms
requires the wavefunction to be totally symmetric under permutations of the
$N$ coordinates.  The time evolution of the wavefunction is given by the
Schr\"odinger equation:
\begin{eqnarray}
i \hbar {\partial \over \partial t} \psi 
&=& \sum_{i=1}^N \left( -{\hbar^2 \over 2m} \nabla_i^2 
\;+\; V(\vec r_i) \right) \psi
\nonumber 
\\ 
&& \;+\; \sum_{i < j} U(\vec r_i - \vec r_j) \psi.
\end{eqnarray}
This formulation of the problem is mathematically awkward if the number of
atoms $N$ is very large and if the effects of the 
interatomic interactions are important.

\subsection{Quantum Field Theory Formulation}

There is an alternative formulation of this
many-body quantum mechanics problem using the language of quantum field
theory.  This formulation is mathematically equivalent,
but it is more convenient if $N$ is very large and if
interatomic interactions are important.  In the quantum field theory
formulation, the basic ingredient is a quantum field operator 
$\Psi(\vec r,t)$, which obeys the following commutation relations:
\begin{eqnarray}
\left[ \Psi (\vec r, t), \Psi (\vec r', t) \right] &=& 0,
\\
\left[ \Psi (\vec r, t), \Psi^\dagger (\vec r', t) \right] &=& 
	\delta^3 (\vec r - \vec r').
\end{eqnarray}
The interpretation of these commutation relations is that the operator
$\Psi(\vec r, t)$ creates an atom at the point $\vec r$ and time $t$ while
$\Psi^\dagger(\vec r, t)$ annihilates an atom.  They also implement the
constraint that the atoms are identical bosons.  The time evolution of the
quantum field is described by the equation
\begin{eqnarray}
&& i \hbar {\partial \over \partial t} \Psi(\vec r,t) \;=\;
\left[ - {\hbar^2 \over 2m} \nabla^2 + V(\vec r) \right] \Psi(\vec r,t)
\nonumber
\\
&& \;+\; \left[ \int d^3 \vec r^\prime \; \Psi^\dagger \Psi(\vec r^\prime,t) 
	U(\vec r^\prime - \vec r) \right] \Psi (\vec r, t).
\label{qfe}
\end{eqnarray}
One can define a ``number operator'',
\begin{equation}
{\cal N} \;=\; \int d^3 r \; \Psi^\dagger \Psi (\vec r, t),
\end{equation}
which is time independent as a consequence of (\ref{qfe}).  
The eigenvalues of ${\cal N}$ can be interpreted as the number of atoms.  
A state containing precisely $N$ atoms is
therefore represented in the quantum field theory by a state $|X_N \rangle$
that is an eigenstate of $\cal N$:
\begin{equation}
{\cal N} | X_N \rangle = N | X_N \rangle.
\end{equation}
It is also convenient to define a ``vacuum state'' $|0\rangle$ containing 
$N=0$ atoms by the condition 
$\Psi (\vec r, t=0) | 0 \rangle = 0$ for all $\vec r$.

The quantum field theory formulation described above is completely equivalent
to the quantum mechanics formulation.  There is a one-to-one
correspondence between Schr\"odinger wavefunctions for $N$ particles 
and $N$-particle states in the quantum field theory.
The Schr\"odinger wavefunction 
$\psi (\vec r_1, \dots , \vec r_N, t)$ that corresponds to an $N$-particle 
state $| X_N \rangle$ can be expressed as a
particular vacuum-to-$|X_N \rangle$ matrix element:
\begin{eqnarray}
&& \psi (\vec r_1, \vec r_2, \dots , \vec r_N, t) 
\nonumber
\\
&& \; \equiv \;
\langle 0 | \Psi (\vec r_1, t) \Psi (\vec r_2, t) 
	\cdots \Psi (\vec r_N,t) | X_N \rangle.
\end{eqnarray}
One of the advantages of the quantum field theory formulation is that $N$
enters into the problem as an eigenvalue rather than as the number of
arguments of the wavefunction.

\subsection{Low-energy approximation}

The evolution equation (\ref{qfe}) 
for the quantum field operator $\Psi (\vec r, t)$ is nonlocal in space.  
However, at the extremely low energies relevant to Bose-Einstein
condensation, it can be replaced by a local equation.  The reason for this is
that at such low energies, the interactions due to the interatomic potential
$U(\vec r_1 - \vec r_2)$ can be described by a single number, the S-wave
scattering length $a$.  The low energy interactions will be unaffected if we
replace $U(\vec r_1 - \vec r_2)$ by any other potential that gives the same
S-wave scattering length.  In particular, we can replace $U(\vec r_1 - {\bf
r}_2)$ by the local potential 
$(4 \pi \hbar^2 a/m) \delta^3 (\vec r_1 -\vec r_2)$.  
The resulting quantum field equation for low energy atoms is
\begin{eqnarray}
i \hbar {\partial \over \partial t} \Psi \;=\;
\left[ - {\hbar^2 \over 2m} \nabla^2 \;+\; V(\vec r) \right] \Psi
\;+\; {4 \pi \hbar^2 a \over m} \Psi^\dagger \Psi \Psi.
\label{qfe-local}
\end{eqnarray}

\subsection{Thermodynamic Limit}

Thus far, we have been considering quantum field equations that are
appropriate for describing a fixed number $N$ of atoms.  If $N$ is extremely
large, it is mathematically more convenient to consider states with an
indefinite number of particles but for which the average value is $N$.  This
can be accomplished by adding a chemical potential term to (\ref{qfe-local}):
\begin{eqnarray}
i \hbar {\partial \over \partial t} \Psi 
&=& \left[ - {\hbar^2 \over 2m} \nabla^2 
	\;+\; V(\vec r) \;-\; \mu \right] \Psi
\nonumber
\\
&& \;+\; {4 \pi \hbar^2 a \over m} \Psi^\dagger \Psi \Psi.
\label{mbqfe}
\end{eqnarray}
The system described by this quantum field equation has a ground state 
$| \Omega \rangle$.  The chemical potential $\mu$ is to be adjusted 
so that the expectation value of the number operator $\cal N$ 
in this state is equal to $N$:
\begin{equation}
\int d^3 r \langle \Omega | \Psi^\dagger \Psi | \Omega \rangle
\;=\; N.
\end{equation}
We expect the RMS deviations of the operator $\cal N$ to scale like
$\sqrt{N}$, which is negligible compared to $N$ if $N$ is sufficiently large.

The quantum field equation (\ref{mbqfe}) 
describes a system consisting of a large number of low
energy atoms interacting through S-wave scattering.  The lagrangian that
summarizes this  quantum field theory is
\begin{eqnarray}
{\cal L} & = & {i \over 2} (\psi^* \dot \psi - \dot \psi^* \psi)
- {1 \over 2m} \nabla \psi^* \cdot \nabla \psi 
\nonumber
\\
&& \;+\; [ \mu - V (\vec r)] \psi^* \psi - {2 \pi a \over m}
(\psi^* \psi)^2,
\label{L1}
\end{eqnarray}
where we have set $\hbar = 1$.  This is a nonrelativistic field theory with a
4-point interaction.  It has a $U(1)$ internal symmetry: 
$\psi \to e^{i \theta} \psi$.

\subsection{Condensates with multiple spin states}

The quantum field theory with lagrangian (\ref{L1})
describes a system consisting of identical 
atoms, all of which  are in the same hyperfine spin state.  
However, it is also possible
to have several spin states of an atom coexisting in a magnetic trap.
The Colorado group has produced condensates containing a mixture of the 
the $|2, +2 \rangle$ and $| 1, -1 \rangle$ spin states of $^{87}$Rb
\cite{JILA-mixed}.  Such a
system would be described by a quantum field theory with two complex fields
$\psi_1$ and $\psi_2$.  The lagrangian includes the interaction terms
\begin{eqnarray}
{\cal L}_{\rm int} \;=\; -g_1 (\psi_1^* \psi_1)^2 
	- g_2 (\psi_2^* \psi_2)^2 
\nonumber
\\
\;-\; g_{12} (\psi_1^* \psi_1)( \psi_2^* \psi_2).
\end{eqnarray}
The three interaction terms correspond to the S-wave scattering of 
$| 1, -1 \rangle$ states, of $|2, +2 \rangle$ states, and
of $| 1, -1 \rangle$ states from $|2, +2 \rangle$ states.

In an optical trap, it is possible to confine all the spin states of a given
hyperfine multiplet.  The MIT group has produced condensates
containing the $|1, +1 \rangle$, $|1, 0 \rangle$, and $|1, -1 \rangle$ 
states of $^{23}$Na \cite{MIT-mult}.  
Such a system is described by a multiplet
$\Psi$ of complex fields, where $\Psi^t = (\psi_{+1}, \psi_0, \psi_{-1})$.
The interaction lagrangian includes two terms:
\begin{equation}
{\cal L}_{\rm int} = -g_0 (\Psi^\dagger \Psi)^2 - g_1 (\Psi^\dagger {\bf S}
\Psi)^2,
\label{L-mult}
\end{equation}
where ${\bf S}$ is the vector of $ 3 \times 3$ spin matrices.  
This lagrangian has a nonabelian $SO(3)$ internal symmetry.
The interaction terms in (\ref{L-mult}) include all the S-wave interactions 
that respect this symmetry.

%%%%%%%%%%%%%%%%%%%%%%%%%%%%%%%%%%%%%%%%%%%%%%%%%%%%%%%%%%%%%%%%%
\section{Equilibrium Properties}
%%%%%%%%%%%%%%%%%%%%%%%%%%%%%%%%%%%%%%%%%%%%%%%%%%%%%%%%%%%%%%%%%

There are several equilibrium properties of a trapped gas of atoms 
that can be studied experimentally.  We will contrast them with
the equilibrium properties of a homogeneous Bose gas.

\subsection{Homogeneous Bose gas}

The thermodynamic properties of a homogeneous Bose gas are determined by the
free energy density.  For a gas at zero temperature, the semiclassical
expansion parameter is $\sqrt { \rho a^3}$.  Thus if the gas is sufficiently
dilute, the system is accurately described by the classical or mean-field
approximation in which the quantum field $\psi(\vec r, t)$ is replaced by its 
ground-state expectation value: 
$\phi = \langle \Omega | \psi(\vec r, t) | \Omega \rangle$.
The mean field $\phi$ is also called the condensate.  
The number density and the free energy density are
\begin{eqnarray}
\rho &=& | \phi |^2,
\\
{\cal F} &=& - \mu | \phi |^2 + {2 \pi a \over m} | \phi |^4 .
\end{eqnarray}
Since the chemical potential $\mu$ is positive, 
the minimum of $\cal F$ occurs at a nonzero value of
$| \phi |$.  There is a continuous set of degenerate minima 
$\phi = \sqrt \rho e^{i \theta}$ parameterized by an angle $\theta$.  
The nonzero value of the
condensate spontaneously breaks the $U(1)$ symmetry.  Solving for the minimum,
we find that the chemical potential and the free energy density are
\begin{eqnarray}
\mu &=& {4 \pi a \over m} \rho,
\\
{\cal F} &=& - {2 \pi a \over m} \rho.
\end{eqnarray}
All the other thermodynamic variables at $T=0$ can be determined 
from $\mu$ and $\cal F$.  
For example, the energy density is $\cal E = \cal F + \mu \rho$ and
the pressure is $P = - \cal F$.

As the temperature is increased from $T=0$ with $\rho$ fixed, the chemical
potential $\mu$ decreases and the condensate $\phi$
decreases.  At some critical temperature $T_c$, $\mu$ vanishes.  
Above $T_c$,
$\mu$ is positive and the condensate vanishes:  $\phi = 0$.
The $U(1)$ symmetry is therefore manifest above $T_c$.  
The phase transition from the symmetric phase with
manifest $U(1)$ symmetry to the spontaneously broken phase as $T$ decreases
through $T_c$ is Bose-Einstein condensation.  
It is a second-order phase transition.  The mean-field approximation
predicts a first order transition and therefore fails to provide an 
adequate description of the system in the critical region near $T_c$.
Careful renormalization group studies of this transition have recently been
carried out by Andersen and Strickland \cite{Andersen-Strickland}.

In summary, the equilibrium properties of a homogeneous Bose gas depend on the
number density $\rho$ and the temperature $T$.  The properties of interest
include the condensate $\phi$, the energy density $\cal E$,
and the pressure $P$.

\subsection{Trapped Bose Gas}

A Bose gas of $N$ atoms trapped in a potential $V(\vec r)$ is a nonhomogeneous
system.  The ground state $| \Omega \rangle$ of this system can still be 
described by the mean-field approximation, provided that the peak value of 
$\sqrt{\rho (\vec r) a^3}$ is sufficiently small.  
In this approximation,  the quantum field $\psi$ is replaced by its 
ground-state expectation value: 
\begin{equation}
\phi(\vec r) \;=\; \langle \Omega | \psi(\vec r, t) | \Omega \rangle.
\end{equation}
The condensate $\phi(\vec r)$ satisfies the Gross-Pitaevskii equation:
\begin{equation}
-{1 \over 2m} \nabla^2 \phi + [ V ( \vec r) - \mu ] \phi 
+ {4 \pi a \over m} | \phi |^2 \phi = 0.
\label{GP}
\end{equation}
The number density is $\rho ( \vec r) = | \phi ( \vec r ) |^2$,
and the chemical potential $\mu$  in (\ref{GP}) must be adjusted so that 
$\int d^3 r \rho ( \vec r ) = N$.
In the ground state, almost all $N$ atoms are in the same quantum state.
The condensate $\phi( \vec r )$ can be interpreted as the 
Schr\"odinger wavefunction of that quantum state.

There are two limits in which the Gross-Pitaevskii equation (\ref{GP}) 
can be solved analytically.  In the weak-interaction limit, 
the condensate is determined by a balance between
kinetic and potential energy.  
If the interaction term in (\ref{GP}) is neglected, 
the Gross-Pitaevskii equation
reduces to the Schr\"odinger equation with energy eigenvalue $\mu$.  For an
isotropic harmonic potential $V(\vec r) = {1 \over 2} m \omega^2 r^2$, the
solution is the familiar Gaussian ground state wavefunction.  The number
density profile is
\begin{equation}
\rho ( \vec r ) \;=\; 
N \left({\pi \over m \omega} \right)^{1/2} e^{-m \omega r^2}.
\end{equation}

The other limit in which the Gross-Pitaevskii equation can be solved
analytically is the strong-interaction limit.  The condensate in this limit is
determined by a balance between the potential and interaction energies.
Neglecting the kinetic term in  (\ref{GP}), 
the Gross-Pitaevskii equation reduces to
an algebraic equation.  For an isotropic harmonic potential, the number
density profile is
\begin{eqnarray}
\rho ( \vec r) 
&=& {15 N \over 8 \pi R^5} \, (R^2 - r^2), 
	\qquad r < R,
\nonumber
\\
&=&  0, \qquad \qquad \qquad \qquad \ r > R. 
\end{eqnarray}
where $R = ( 15 Na/2 m^2w^2)^{1/5}$ is the radius of the condensate.  
In contrast to the weak interaction limit, 
the condensate has a sharper edge and its size
is larger by a factor of order $(Na \sqrt {m \omega})^{1/5}$.

The {\it density profile} 
$\rho(\vec r)$ of a condensate can be measured
by illuminating it with a laser beam.  The attenuation of the 
laser light increases with the column density, which is the integral of
$\rho(\vec r)$ along the path of the laser beam.
Measurements of the density profiles of the condensate produced in present
experiments have shown that they are very close to the strong interaction
limit.  In typical experiments, the radius of the condensate is
larger than that of ground-state wavefunction of a single atom 
by a factor of 5 to 10.
One should not think of these systems as ideal gases
of almost noninteracting atoms.  The
interactions between atoms have a dramatic effect on the condensate.

Given the density profile of a condensate at nonzero temperature,
one can determine the {\it condensate fraction} $N_0/N$, where $N_0$ 
is the number of atoms in the condensate.  One simply fits the observed 
density profile $\rho(\vec r)$ to the sum of the 
density profiles of a condensate and a thermal cloud.  
The integral of the condensate density 
gives the number $N_0$.  The condensate fraction $N_0/N$ decreases from 
nearly 100\% to 0 as the temperature is increased 
from zero to the critical temperature.

Another equilibrium property that can be measured is the 
{\it release energy}, which is the sum of the kinetic and interaction 
energies of the atoms:  
\begin{eqnarray}
E_R \;=\; \int d^3r
\left( {1 \over 2m} \nabla \phi^* \cdot \nabla \phi 
\;+\; {2 \pi a \over m} |\phi|^4 \right).
\label{E-R}
\end{eqnarray}
This can be measured by suddenly turning off the 
trapping potential $V(\vec r)$.  This sets the potential energy to 0,
and the remaining energy is given by (\ref{E-R}).
Since the atoms are no longer confined by the 
potential, they will drift away.  As the cloud of atoms
expands, its density decreases and all its interaction energy is 
converted into kinetic energy.
By measuring the density profile as a function of time,
one can deduce the sum of the final kinetic energies, 
and this is equal to the release energy $E_R$.
The evolution of the Bose-Einstein condensate after switching off the
trapping potential has been recently studied using quantum  
field theory methods by Nakamura and Yamanaka \cite{Nakamura-Yamanaka}.

In summary, the measurable equilibrium properties of a trapped Bose gas
of atoms are rather different from those of a homogeneous Bose gas.
They include the density profile $\rho(\vec r)$, the condensate fraction 
$N_0/N$, and the release energy $E_R$.

%%%%%%%%%%%%%%%%%%%%%%%%%%%%%%%%%%%%%%%%%%%%%%%%%%%%%%%%%%%%%%%%%
\section{Collective Excitations}
%%%%%%%%%%%%%%%%%%%%%%%%%%%%%%%%%%%%%%%%%%%%%%%%%%%%%%%%%%%%%%%%%

Some of the most beautiful experiments on Bose-Einstein condensates  
involve studying their collective excitations.  We will contrast 
the collective excitations of
a trapped gas with those of a homogeneous Bose gas.

\subsection{Homogeneous Bose gas}

The collective excitations of a homogeneous Bose gas at $T=0$
were first studied by Bogoliubov.  He found that the quasiparticles
in the gas have a dispersion relation of the form
\begin{equation}
\omega(k) \;=\; { k \sqrt{k^2 + 1/\xi^2} \over 2 m}.
\end{equation}
where $\xi = (16 \pi a \rho)^{-1}$ is the coherence length.
For large $k$, this approaches the energy $k^2/2m$ of a free atom.
For small $k$, the dispersion relation is linear:
$\omega(k) \to k/(2 m \xi)$.
This should not be a surprise, because our quantum field theory has a 
$U(1)$ symmetry that is spontaneously broken.  There must therefore be a 
Goldstone mode. The Bogoliubov dispersion relation interpolates between 
that of the Goldstone mode at small $k$ and that of a 
free particle at large $k$, with the crossover occurring at $k$ of order
$1/\xi$.

At nonzero temperatures below $T_c$, the collective excitations are 
more complicated, because the condensate coexists with a thermal gas of atoms.
One can think of this system as consisting of two interpenetrating fluids,
with the thermal gas being a normal fluid and the condensate being 
a superfluid.  The collective excitations involve
coupled oscillations of the condensate and the thermal gas.
A further complication is that the oscillations damp out with time,
as energy in the collective modes is transferred 
to other modes of the system.  Thus the basic properties of the 
collective excitations in the homogeneous gas are the
dependence of the oscillation frequencies and their damping rates
on the temperature and density.

\subsection{Trapped Bose Gas}

For a Bose gas of $N$ atoms trapped in a potential $V(\vec r)$,
the collective excitations at zero temperature can 
be described by the mean-field approximation, provided that the 
peak value of $\sqrt { \rho (\vec r) a^3}$ is sufficiently small.  
In the state $| X \rangle$ in the quantum field theory that 
corresponds to a collective excitation, the expectation value 
of the quantum field $\psi(\vec r,t)$ changes with time.  The condensate 
is therefore time dependent:
\begin{equation}
\phi(\vec r,t) \;=\; \langle X | \psi(\vec r,t) | X \rangle
\end{equation}
In the mean-field approximation, it satisfies the 
time-dependent Gross-Pitaevskii equation:
\begin{equation}
i {\partial \over \partial t} \phi
\;=\; -{1 \over 2m} \nabla^2 \phi + [ V ( \vec r) - \mu ] \phi 
+ {4 \pi a \over m} | \phi |^2 \phi.
\label{GP-t}
\end{equation}
Almost all the atoms are in the same time-dependent quantum state,
and the condensate $\phi(\vec r,t)$ can be interpreted as the 
Schr\"odinger wavefunction for that quantum state.

For an isotropic harmonic potential
$V(\vec r) = {1 \over 2} m \omega^2 r^2$, the frequencies for 
small-amplitude oscillations can be calculated analytically
in two limits.  In the weak-interaction limit, 
(\ref{GP-t}) reduces to the Schr\"odinger equation for the harmonic oscillator.
The normal modes are labelled by the principal quantum number 
$n = 0,1,2,...$ and by the angular momentum quantum numbers $\ell$ and $m$.
The oscillation frequencies are
\begin{equation}
\omega_{n \ell} \;=\; (2 n + \ell) \; \omega.
\end{equation}
The other limit in which one can obtain analytic expression for
the frequencies  is the strong-interaction limit.  
The spectrum in this limit was first obtained by Stringari \cite{Stringari}:
\begin{equation}
\omega_{n \ell} \;=\; 
\sqrt{2 n^2 + 3 n + (2 n + 1) \ell} \; \omega.
\end{equation}

There are several normal modes of a Bose-Einstein condensate 
that can be excited by temporarily changing the trapping potential.
Both the Colorado group \cite{JILA-osc} and the MIT group \cite{MIT-osc} 
have measured the oscillation frequencies and damping rates
of condensates excited in this way. For condensates 
containing large numbers of atoms, measurements of the oscillation 
frequencies have verified that these systems are close 
to the strong interaction limit.  
In some cases, the oscillation frequencies have been 
measured with precisions of better than a percent.
This is approaching the accuracy necessary to see the
corrections to the mean-field approximations that arise
from quantum field fluctuations.

It is also possible to study the propagation of sound in a condensate. 
By using laser beams to modify the trapping potential,
the MIT group \cite{MIT-sound}
has created local density perturbations in a long 
cigar-shaped condensate.  When the laser beams are turned off,
the density perturbations propagate along the length of the condensate.
The speed of propagation provides a measure of the speed of sound
in the condensate.

%%%%%%%%%%%%%%%%%%%%%%%%%%%%%%%%%%%%%%%%%%%%%%%%%%%%%%%%%%%%%%%%%
\section{Conclusions}
%%%%%%%%%%%%%%%%%%%%%%%%%%%%%%%%%%%%%%%%%%%%%%%%%%%%%%%%%%%%%%%%%

In summary, the Bose-Einstein condensation of atoms is  
described by a thermal quantum field theory.
The interactions between the atoms have dramatic effects on the condensate,
and this makes the problem interesting and nontrivial.
The interactions are local and have a simple structure, 
so the lagrangian that describes the system
is known.  The most remarkable feature is that there 
is experimental control over almost every aspect of the system.
Bose-Einstein condensation has been achieved with several different atoms, 
and for each atom there are several choices of spin states.
The trapping potential $V$ can be controlled experimentally,
not only as a function of $\vec r$ but also as a function of $t$.
The number of atoms in the trap and the temperature of the atoms 
can also be controlled.  Even the interaction strength of the atoms 
can be controlled.  The Colorado group \cite{JILA-change}
has demonstrated that by using laser pulses, all the atoms in a condensate 
consisting of the $|1,-1 \rangle$ state of $^{87}$Rb can be changed 
instantaneously into the $|2,+2 \rangle$ state.  Since the $|2,+2 \rangle$
state has a different scattering length, this amounts to an 
instantaneous change in the scattering length.  
An even more dramatic development has come from the MIT group
\cite{MIT-change}, which has demonstrated that, by using magnetic fields
to adjust the atoms in an optical trap to a Feshbach resonance, 
the scattering amplitude can be adjusted to any desired
value, even $+ \infty$.   In addition to marvelous experimental control,
there are also beautiful experimental probes that can be used to study
the equilibrium properties and the collective excitations of Bose-Einstein
condensates.  Thus this system provides an ideal testing ground for the
methods of thermal field theory. 

I hope this paper has whetted the appetites of thermal field theorists
to learn more about the Bose-Einstein condensation of atoms.
Fortunately, there are several recent review articles that provide a good 
introduction to this subject. The Nordita lecture notes
by Pethick and Smith \cite{Pethick-Smith} provide a good overview
of the physics of Bose-Einstein condensation.  A review
article by Shi and Griffin \cite{Shi-Griffin} summarizes the 
current state of understanding of the homogeneous Bose gas.
The theory of Bose-Einstein condensation in a trapped Bose gas
has been reviewed by Dalfovo et al. \cite{D-G-P-S}. 

Most of the important theoretical developments in this field 
thus far have come from physicists in the condensed matter physics
community.  The three review articles mentioned above include rather 
exhaustive references to this work.  
For a sampling of interesting work from a condensed matter perspective,
I refer the reader to some papers by my colleague T.-L. Ho \cite{Ho}.
Researchers coming from a 
background in quantum field theory might find some of the following
papers more accessible.  There is a beautiful paper by Liu \cite{Liu}
that explains how effective-field-theory methods can be used to 
construct model-independent lagrangians that describe the low-energy
collective excitations of a Bose-condensed gas at nonzero temperature.
Haugset, Haugerud, and Ravndal \cite{H-H-R} have calculated the 
one-loop effective potential for a Bose gas at nonzero temperature
using functional methods.  Nieto and I \cite{Braaten-Nieto:3}
have calculated the free energy of a Bose gas at zero temperature 
to two-loop accuracy.  These papers all involve applications of 
quantum field theory to the homogenous Bose gas.  However quantum 
field theory methods can also be useful for the experimentally 
achievable case of a trapped gas.  
For example, Nieto, Andersen, and I 
\cite{Braaten-Nieto:2} have shown how semiclassical 
corrections to the properties of a Bose-Einstein condensate
can be taken into account by adding local correction terms to the 
classical field equations of the mean-field approximation.
These methods greatly simplify the study of semiclassical corrections.

%%%%%%%%%%%%%%%%%%%%%%%%%%%%%%%%%%%%%%%%%%%%%%%%%%%%%%%%%%%%%%%%%
% References:
%%%%%%%%%%%%%%%%%%%%%%%%%%%%%%%%%%%%%%%%%%%%%%%%%%%%%%%%%%%%%%%%%

\acknowledgments
This work was supported in part by the U.~S. Department of Energy,
Division of High Energy Physics, under Grant DE-FG02-91-ER40690
and by a Faculty Development Grant from the Physics Department of The 
Ohio State University.

\end{document}